\documentclass[prl,aps,floatfix,preprint]{revtex4-1}

\usepackage{amsmath,amssymb}
\usepackage{graphicx}
\usepackage{psfrag}
\usepackage{graphicx}
\usepackage{color}
\usepackage[normalem]{ulem}
\usepackage{lipsum}

\def\beq{\begin{equation}}
\def\eeq{\end{equation}}
\def\bea{\begin{eqnarray}}
\def\eea{\end{eqnarray}}

\begin{document}

\title{Casimir stresses in  active nematic films}
\author{Abhik Basu$^{1,3}$, Jean-Francois Joanny$^{2,4}$, Frank J\"ulicher$^3$,
Jacques Prost$^{2}$} \affiliation{{$^1$Condensed
Matter Physics Division, Saha Institute of Nuclear Physics, 1/AF,
Bidhannagar, Calcutta 700 064, India, \\
$^2$Physicochimie Curie (CNRS-UMR168), Institut Curie, Section de
Recherche, 26 rue
d'Ulm 75248 Paris Cedex 05 France, \\
 $^3$Max-Planck-Institut f\"ur
Physik komplexer Systeme, N\"othnitzerstr. 38, 01187 Dresden,
Germany, \\
$^4$E.S.P.C.I, 10 rue Vauquelin, 75231 Paris Cedex 05, France} }

\date{\today}

\begin{abstract}
We calculate the Casimir stresses in a thin
layer of active fluid  with nematic order. By using a stochastic 
hydrodynamic approach
for an active fluid layer of finite thickness $L$, we generalize the Casimir 
stress for nematic liquid crystals in
thermal equilibrium to active systems. We show that the active Casimir stress 
differs  significantly from its
equilibrium counterpart.  For contractile activity, the active Casimir 
stress, although attractive like its equilibrium counterpart, diverges
logarithmically as $L$ approaches a threshold of  the
spontaneous flow instability from below. In contrast, for small extensile activity, it is repulsive, has no divergence at any $L$ and has a scaling with $L$ different from its equilibrium counterpart.

\end{abstract}

\maketitle

\section{Introduction}

It is well-known that although the zero-point energy of the
electromagnetic field inside a cavity bounded by conducting walls is
formally diverging, its variation upon displacements of the
boundaries remains finite. It corresponds to a weak but measurable
attractive force, known as the Casimir force \cite{casimir}. For
example, in the case of two parallel conducting plates at a distance
$L$, the attractive Casimir force per unit area, or the Casimir stress is
given by $C_F=-\frac{\pi^2}{240}\frac{h 2\pi
c}{L^4}$~\cite{casimir}. It is of purely quantum origin.

Subsequently, thermal analogs of the Casimir stress associated to various
fluctuating
fields at a finite temperature $T$
have been studied. In  nematic liquid crystals confined between
two parallel plates, the thermal fluctuations
of the director field that describes the nematic order, play the role of
the electromagnetic fluctuations in the electromagnetic
Casimir effect.  In all such classical systems, the boundary conditions on
the relevant fields (e.g., the director field for
nematic liquid crystals) constrain their thermal fluctuations and
lead to a thermal analog of the Casimir stress. For instance,
for a nematic liquid crystal between parallel confining plates separated by a distance $L$ with the
director field rigidly anchored to them, one again obtains an
attractive Casimir stress that varies with the thickness  $L$ of the liquid 
crystal film as
$1/L^3$~\cite{ajdari}.

Studies on non-equilibrium analogs of
thermal Casimir stresses are relatively new.
In Ref.~\cite{casinon1}, Casimir stresses between two
parallel plates due to non-thermal noises are calculated. Further,
embedding objects or inclusions in a correlated 
fluid are shown to generate
 effective Casimir-like stresses between the inclusions~\cite{casinon2}. There are 
direct biologically relevant examples as well:
more recently, Ref.~\cite{amit1} elucidated the dependence of this 
Casimir-like forces on inclusions in a fluctuating active fluids on active 
noises and hydrodynamic interaction of the inclusion with the boundaries. 
Subsequently, Ref.~\cite{amit2} studied the role of active Casimir effects on 
the deformation dynamics of the cell nucleus and showed the appearance of a 
fluctuation maximum at a critical level of activity, a result in agreement with 
recent experiments~\cite{makhija}. The active fluid models considered by 
Refs.~\cite{amit1,amit2} are effectively one-dimensional and hence do not 
include any soft orientational fluctuations.

In this article, we
calculate the Casimir stress between two parallel
plates confining a layer of an active  nematic fluid with a uniform macroscopic
orientation~\cite{kruse,sriram,sriram1}. The active fluid is driven out of 
equilibrium
by a locally constant supply of energy. Our work directly generalizes thermal 
Casimir
stresses in equilibrium nematics~\cite{ajdari} into the nonequilibrium domain.

The hydrodynamic active fluid model~\cite{kruse,sriram} has been proposed as a 
generic 
coarse-grained
model for a driven orientable fluid with nematic or polar symmetry.
The main feature of the active fluid is the existence of an {\em active
stress} of non-equilibrium origin that describes
the constant consumption of energy by the system, which drives the
system away from equilibrium. Due to its very general nature, the active fluid
model is
able to describe a broad range of phenomena, observed in very different physical
systems and at very different length scales~\cite{kruse,sriram,sriram1}. 
Notable examples include the dynamics
of actin filaments in the cortex of eukaryotic cells or
bird flocks  and bacterial biofilms. In particular, in the case of
actin filament dynamics, the active stress results from the release of
free energy due to the chemical conversion of Adenosine-Triphosphate (ATP) to
Adenosine-Diphosphate (ADP).

In this article, we study Casimir forces using a
stochastically driven coarse-grained hydrodynamic approach for
active fluids~\cite{kruse,sriram,sriram1}, with a nematic order,
described by a unit vector polarization field $p_\alpha,\,\alpha=x,y,z$. The
film is infinite along the $x,y$ plane, but has a finite thickness
$L$ in the $z$-direction. A typical example of ordered active nematic where our
results may apply is
the cortical actin layer in a cell where the orientation of the
actin filaments can have a component parallel to the cell membrane.
It has been recently shown that a liquid contractile active film of thickness
$L$ with polarization either parallel or perpendicular to its
surface has a spontaneous flow instability, above a critical value
of the activity \cite{voituriez,sumithra}. This is the nonequilibrium analog
of the ``Frederiks transition'' in equilibrium classical
nematic liquid crystals. It
is driven by the coupling between the polarization orientation and
the active stress. We here calculate $C$, the {\em active} analog of the
thermal equilibrium Casimir  stress, that we formally define below. 

\section{Active Casimir stress}

 We consider a thin film of active fluid with a fixed thickness $L$ along 
the $z$-direction confined between the planes $z=0$ and $z=L$. 
In the passive case, i.e., without any activity, the Casimir stress $C_{eq}$ is defined as\cite{ajdari}
\begin{equation}
 C_{eq}=\langle \sigma_{zz}^{eq}\rangle|_{z=L}- \langle \sigma_{zz}^{eq}\rangle|_{z=\infty}.\label{ceq}
\end{equation}
Here, $\sigma_{zz}^{eq}$ is the normal component of the equilibrium stress that diverges for all $z$ (or, all $L$); $C_{eq}$ however is finite for any non-zero $L$\cite{ajdari}. Here, $\langle..\rangle$ implies averages of thermal noise ensembles (see below). 
In an active system, we define the Casimir stress $C$ as 
\begin{equation}
 C=\langle\sigma^{tot}_{zz}\rangle|_{z=L}-\langle\sigma^{tot}_{zz}
\rangle|_{\Delta\mu=0,z=L} - \left[\langle\sigma^{tot}_{zz}\rangle|_{z=L}-\langle\sigma^{tot}_{zz}
\rangle|_{\Delta\mu=0,z=L}\right]_{K\rightarrow \infty}, \label{casidefn11}
\end{equation}
where $K$ is the Frank elastic constant of the nematics (assuming a one Frank constant description). Here, $\sigma^{tot}_{zz}$ is the normal component of the total stress in an active fluid and $\Delta\mu$ is the activity parameter that parametrize the free energy release in the chemical conversion of ATP to ADP.
Here,
$\sigma^{tot}_{zz}|_{\Delta\mu=0}=\sigma^{eq}_{zz}$, the normal component of 
the equilibrium stress. Note that the last term in  
(\ref{casidefn11}) in the limit $K\rightarrow \infty$ represents the stresses $\langle\sigma ^{tot}_{zz}\rangle$ 
and $\langle \sigma^{eq}_{zz}\rangle$ in the absence of any orientation 
fluctuations which are independent of layer thickness $L$. By using stochastic hydrodynamic 
descriptions for orientationally ordered active fluids, we show below that 
(\ref{casidefn11}) reduces to
\begin{equation}
 C= -\frac{K}{2}\langle (\partial_z 
p_i)^2\rangle|_{z=L} + \frac{K}{2}\langle (\partial_z 
p_i)^2\rangle|_{\Delta\mu=0,z=L}.\label{casidefn111}
\end{equation}
 The quantity $C$ is difficult to measure directly. However changes of $C$ due to changes in $L$  can in principle be measured.


When the thickness $L$ of a contractile
active fluid layer  approaches the critical
thickness $L_c$ for the  spontaneous flow instability from below~\cite{voituriez}, we  show that $C$ remains attractive, scales with $L$ in a way same as its equilibrium counterpart, but diverges logarithmically as $L$ approaches $L_c$ from below.  We also calculate $C$ for extensile activity, and contrast it with the active Casimir stress for the contractile case: in this case,  $C$ is found be repulsive, has no divergence at any finite $L$, and scales with $L$ {\em differently} from the equilibrium result.

\section{Stead state stresses in a fluctuating active fluid}

We consider an incompressible viscous active fluid film with nematic 
order. Our
analysis below closely follows the physical discussion of
Ref.~\cite{bead}, where the diffusion coefficient of a test particle
immersed in an active fluid with nematic order is calculated. The force
balance in an incompressible active fluid is given by
\begin{equation}
\partial_\beta (\tilde \sigma_{\alpha\beta}
+\sigma^a_{\alpha\beta}-
P\delta_{\alpha\beta}+\sigma^e_{\alpha\beta})=0,\label{fullstress}
\end{equation}
 where fluid inertia is neglected~\cite{kruse,basu}.
Here, $\tilde
\sigma_{\alpha\beta}$ denotes the traceless part of the symmetric
deviatoric stress and the antisymmetric deviatoric stress is given
by
\begin{equation}
\sigma^a_{\alpha\beta}=\frac{1}{2}(p_\alpha h_\beta-p_\beta
h_\alpha). \label{antisymm}
\end{equation}
Here $h_\alpha=-\delta F/\delta p_\alpha$ is the orientational field
conjugate to the  nematic director $p_\alpha$, where $F=\int d^3r f$
denotes the  nematic director free energy with a free energy density 
$f$.  Furthermore, $P$ denotes the hydrostatic pressure.
 Note
that in a  nematic system the equilibrium stress can have anisotropies
described by the Ericksen stress
\begin{equation}
\sigma^e_{\alpha\beta} =-\frac{\partial f}{\partial(\partial_\beta
p_\gamma)}\partial_\alpha p_\gamma \quad .\label{erick}
\end{equation}
Here, $\alpha,\beta=x,y,z$. The total normal stress is thus given by
\begin{equation}
 \sigma^{tot}_{\alpha\beta}=\tilde \sigma_{\alpha\beta}
+\sigma^a_{\alpha\beta}-
P\delta_{\alpha\beta}+\sigma^e_{\alpha\beta}; \label{fullstress1}
\end{equation}
see Eq.~(\ref{fullstress}) above.

In the following, we impose  for simplicity a constant amplitude of the 
 nematic director
$p_\gamma p_\gamma=1$. The constitutive
equations of  a single-component active fluid then read~\cite{basu}
\begin{eqnarray}
\left\{ \tilde \sigma_{\alpha\beta}+\zeta\Delta\mu q_{\alpha\beta}
+\frac{\nu_1}{2}(p_\alpha h_\beta +p_\beta
h_\alpha-\frac{2}{3}p_\gamma h_\gamma\delta_{\alpha\beta})
\right\}&=&2\eta v_{\alpha\beta}+\xi^{{\sigma}}_{\alpha\beta},\label{stresseq} \\
\frac{D}{D t}p_\alpha = \frac{1}{\gamma_1}h_\alpha -\nu_1 p_\beta
\tilde v_{\alpha\beta}&+&\xi_{\perp \alpha} \label{eq:pdyn}
\end{eqnarray}
where $q_{\alpha\beta}=(p_\alpha p_\beta-\frac{1}{3}\delta_{\alpha\beta})$
is the nematic tensor. The symmetric velocity
gradient tensor is 
$\tilde v_{\alpha\beta} = (\partial_{\alpha} v_{\beta}+\partial_{\beta} 
v_{\alpha})/2$, where 
$v_{\alpha}$ 
is the three-dimensional velocity field of the active fluid ($\alpha=x,y,z$). 
The shear
viscosity is denoted by $\eta$, $\gamma_1$ is the
rotational viscosity and $\nu_1$ the flow alignment parameter which
is a number  of order one.   
Functions $\xi_{\alpha\beta}^\sigma$ and $\xi_{\perp\alpha}$ are stochastic 
noises, which we assume 
to be thermal noises of zero-mean and variances given by
\begin{eqnarray}
\langle \xi^{\sigma}_{\alpha\beta}(t,{\bf x})
\xi^{\sigma}_{\gamma\delta}(t',{\bf x}') \rangle &=& 2k_B T \eta
\left [
(\delta_{\alpha\gamma}\delta_{\beta\delta}+\delta_{\alpha\delta}\delta_{\beta\gamma}-
\frac{2}{3}\delta_{\alpha\beta} \delta_{\gamma\delta}\right ]
\delta(t-t')\delta({\bf x}-{\bf x}'),\label{noiseI} \\
\langle \xi_{\perp \alpha} (t,{\bf x}) \xi_{\perp \beta}(t',{\bf
x}') \rangle &=& 2\frac{k_B T}{\gamma_1}[
\delta_{\alpha\beta}-p_\alpha p_\beta ]\delta(t-t')\delta({\bf
x}-{\bf x}').\label{noiseII}
\end{eqnarray}
where $k_B$ is Boltzmann constant and $T$ denotes temperature.   Notice 
that the noises $\xi_{\perp \alpha} (t,{\bf x})$ are {\em multiplicative} in 
nature (see noise variance (\ref{noiseII})). However, since we are interested in 
a linearized description about uniform ordered states (see below), the 
multiplicative nature of these noises do not affect us. Furthermore, we do not 
consider 
any athermal or active noises for simplicity.
 We consider an incompressible system imposed by the constraint 
$\partial_\alpha v_\alpha=0$.

The pressure $P$ plays the role of 
 a Lagrange multiplier used to impose 
the incompressibility constraint $\partial_\alpha v_\alpha=0$.   The 
incompressibility leads to the following equation for $P$:
\begin{equation}
\nabla^2 P=-\frac{\nu_1}{2}\partial_\alpha\partial_\beta (p_\alpha
h_\beta + p_\beta h_\alpha-\frac{2}{3}{\bf p\cdot
h}\delta_{\alpha\beta})-\zeta\Delta\mu\partial_\alpha\partial_\beta
(p_\alpha p_\beta)+\partial_\alpha\partial_\beta
\sigma^e_{\alpha\beta}+\partial_\alpha\partial_\beta\xi^\sigma_{\alpha\beta}.
\label{pressure1}
\end{equation}

 We consider a film of the active fluid with a fixed 
thickness $L$ along the $z$ direction, confined between the planes $z=0$ and 
$z=L$. We consider a non-flowing reference state together 
with $p_z=1$, which is a steady state solution of (\ref{fullstress}) and 
(\ref{eq:pdyn}). We study small 
fluctuations $\delta {\bf p}=(p_x,p_y,0)$ around this 
state; $\delta p = |\delta {\bf p}|$. We impose boundary conditions $
(p_x, p_y)=0$ and  vanishing shear stress 
at $z=0$ and $z=L$. The total normal stress on the surface at $z=L$, $\langle 
\sigma_{zz}^{tot}\rangle_{z=L}$ should depend on $L$ and also contains a 
constant piece independent of $L$~\cite{ajdari}. 
From the definition of $\sigma_{zz}^{tot}$
\begin{eqnarray}
\langle\sigma_{zz}^{tot}\rangle_{z=L}&=& 
\eta\langle 
\frac{\partial
v_z}{\partial z}\rangle_{z=L} - \zeta\Delta\mu \langle p_z^2\rangle|_{z=L}
-\frac{\nu_1}{3} \langle p_i h_i\rangle_{z=L}
-\frac{\nu_1}{3}\langle p_z
h_z\rangle|_{z=L}\nonumber \\ &+&\langle\sigma^e_{zz}\rangle_{z=L} -\langle
P\rangle_{z=L}.\label{castot}
\end{eqnarray}
Here, $i,j=x,y$ are the coordinates along the film surface. Using, for 
simplicity and analytical
convenience, a single Frank elastic constant $K$ for the nematic liquid
crystals, the Frank free energy density is given by $f=K(\nabla_\alpha 
p_\beta)^2/2$.
Below we evaluate the pressure $P$
 which obeys Eqs~(\ref{pressure1}). The 
remaining terms in (\ref{castot}) are also to be evaluated using the relevant 
equations of motion and then averaging
over the various noise terms. 
The contributions to the stress that are linear in small
fluctuations $\delta {\bf p}$ vanish upon averaging;
therefore, a non-vanishing Casimir stress is obtained from 
contributions to the stress
quadratic in $\delta {\bf p}$ in (\ref{castot}).  It is instructive to analyze 
the different contributions in (\ref{castot}) to $C$ term by term. This will 
allow us to considerably simplify (\ref{castot}) as we will see below.

We first consider the  contribution $\eta \langle \frac{\partial 
v_z}{\partial z}\rangle_{z=L}$ in (\ref{castot}). Using the condition of 
incompressibility ${\boldsymbol\nabla}\cdot {\bf v}=0$, this may be written as 
\begin{equation}
\eta\langle \frac{\partial v_z}{\partial z}\rangle_{z=L} = - \eta
\langle{\boldsymbol \nabla}_\perp \cdot {\bf v}_\perp\rangle_{z=L}
=-\eta {\boldsymbol\nabla}_\perp \cdot\langle {\bf v}_\perp \rangle_{z=L}
=0,\label{term1}
\end{equation}
since, there is no flow on an average. Here,
${\boldsymbol\nabla}_\perp = (\frac{\partial}{\partial
x},\frac{\partial}{\partial y})$ is the two-dimensional gradient
operator and ${\bf v}_\perp= (v_x,v_y)$ is the in-plane component of
the three-dimensional velocity $\bf v$.

Secondly, $\langle p_i h_i\rangle_{z=L}=0$ since $p_i=0$ at $z=L$. Further,
$\langle p_z h_z\rangle_{z=L} = \langle h_\parallel\rangle$, since $p_z=1$ at 
$z=L$. Here, $h_\parallel$ is a Lagrange multiplier, which must be introduced 
to impose $p^2=1$, or to the leading order $p_z=1$ in the geometry that we 
consider. Using $p_z=1$ in Eq.~(\ref{eq:pdyn}) and linearizing around $p_z=1$, 
we obtain
$ h_\parallel\sim  \frac{\partial
v_z}{\partial z}$  at all $z$~\cite{basu}. Using the incompressibility 
condition, $\partial v_z/\partial z= - {\boldsymbol\nabla}_\perp \cdot {\bf 
v}_\perp$. This then gives $\langle h_z\rangle=0$ to the leading order in 
fluctuations.

{ In order to evaluate the form of the pressure $P$, we consider the 
equation for the velocity field $v_\alpha$ that obeys the generalized Stokes 
equation
\begin{equation}
 \eta \nabla^2 v_\alpha = \partial_\alpha P +\zeta\Delta\mu 
\partial_\beta(p_\alpha p_\beta) -\frac{\nu_1}{2}\partial_\beta (p_\beta 
h_\alpha + p_\alpha h_\beta) -\frac{1}{2}\partial_\beta (p_\alpha h_\beta - 
p_\beta h_\alpha) -\partial_\beta\sigma_{\alpha\beta}^e 
-\partial_\beta\xi^\sigma_{\alpha\beta}. \label{genstok11}
\end{equation}
We focus on the in-plane velocity $v_i,\,\alpha=i=x,y$ in (\ref{genstok11}) above.
Now consider the different terms in (\ref{genstok11}) with $\alpha=i$ at $z=L$  and note that 
(i) $p_i=0$ at $z=L$, (ii) in the absence of any mean flow and consistent 
with the in-plane rotational invariance, velocity fluctuations $+v_i$ and 
$-v_i$ should be equally likely in the statistical steady state, i.e.,  the steady 
state average of any function odd in $v_i$ should be zero. This implies that
$\langle v_i\rangle =0 = \langle \partial_z^2 v_i\rangle$ in steady states. 
Similarly, in an oriented state having nematic order with $p_z=1$, fluctuations 
$+p_i$ and $-p_i$ should be equally likely in the steady state, i.e., any 
function odd in $p_i$ must have a vanishing average in the steady state.
Furthermore, since $h_i$ is odd in $p_i$, we must have $\langle 
h_i\rangle =0$ in the steady states. Similarly,
\begin{eqnarray}
 \zeta\Delta\mu \langle \partial_\beta (p_i p_\beta)\rangle|_{z=L}
 =\zeta\Delta\mu \langle \partial_j (p_i p_j)\rangle|_{z=L} + \zeta\Delta\mu 
\langle \partial_z (p_i p_z)\rangle|_{z=L}=\zeta\Delta\mu 
\langle \partial_z (p_i p_z)\rangle|_{z=L}
\end{eqnarray}
vanishes in the steady state due to the inversion symmetry of $p_i$. 
Lastly, we note that
\begin{eqnarray}
 \partial_\beta \sigma_{i\beta}^e|_{z=L}=-\partial_j (\partial_i p_\gamma 
\partial_j p_\gamma)|_{z=L} - \partial_z (\partial_i p_\gamma \partial_z 
p_\gamma)|_{z=L}=-\frac{1}{2}\partial_i (\partial_z p_j)^2|_{z=L},
\end{eqnarray}
where we have used $[(\partial_i p_\gamma)(\partial_z^2 p_\gamma)]_{z=L}=[(\partial_i p_j)(\partial_z^2 p_j)]_{z=L} + [(\partial_i p_z)(\partial_z^2 p_z)]_{z=L}=0$, since $p_j=0$ and $p_z=1$ exactly at $z=L$.
Putting together everything and averaging 
in the steady states, 
we then obtain at $z=L$
\begin{equation}
 \partial_i P=\partial_\beta\sigma^e_{i\beta}=-\frac{K}{2}\partial_i 
(\partial_z 
p_j)^2,
\end{equation}
giving
\begin{equation}
 P=\frac{K}{2}(\partial_z 
p_j)^2 +a_0
\end{equation}
at $z=L$, where $a_0$ is a constant of integration. Then substituting $P$ in 
(\ref{castot})
\begin{equation}
 \langle\sigma_{zz}^{tot}\rangle_{z=L}=-\frac{K}{2}\langle (\partial_z 
p_i)^2\rangle|_{z=L} -\zeta\Delta\mu + a_0 = -\frac{K}{2}\langle (\partial_z 
p_i)^2\rangle|_{z=L} + \tilde a_0, \label{casi1}
\end{equation}
where $\tilde a_0$ is another constant.
 
Notice that the constant $\tilde a_0$, which in general can depend 
upon 
$\Delta\mu$ is actually $\langle\sigma_{zz}^{tot}\rangle$ evaluated in 
the limit $K\rightarrow \infty$ (i.e., with all the orientation fluctuations 
suppressed): $\tilde a_0 = \langle\sigma_{zz}^{tot}\rangle|_{K\rightarrow \infty}$ and is independent of $L$. Similarly in the passive case~\cite{ajdari}
\begin{equation}
 \langle\sigma^{tot}_{zz}
\rangle|_{\Delta\mu=0,z=L}=\langle\sigma_{zz}^{eq}\rangle_{z=L} = -\frac{K}{2}\langle (\partial_z 
p_i)^2\rangle|_{z=L,\Delta\mu =0} + a_0^{eq},
\end{equation}
where $a_0^{eq}$ is a constant that is given by $\langle\sigma_{zz}^{eq}\rangle|_{K\rightarrow\infty}$ and is independent of $L$. 
We are now in a position to formally define active Casimir stress $C$ as
\begin{eqnarray}
  C&=&\langle\sigma^{tot}_{zz}\rangle|_{z=L}-\langle\sigma^{tot}_{zz}
\rangle|_{\Delta\mu=0,z=L}-\tilde a_0+a_0^{eq}\nonumber \\
&=&-\frac{K}{2}\langle (\partial_z 
p_i)^2\rangle|_{z=L} + \frac{K}{2}\langle (\partial_z 
p_i)^2\rangle|_{\Delta\mu=0,z=L} . \label{casidefn}
\end{eqnarray}

}


   We 
show below that $C$ in an ordered active 
nematic layer is
fundamentally different from its equilibrium counterpart, primarily because the 
dynamics of
orientation fluctuations here is very different from its equilibrium 
counterpart.

 We  calculate $C$ for small fluctuations 
around the chosen 
reference state by using the dynamical equations (\ref{fullstress}) and 
(\ref{eq:pdyn}). Since $\langle\sigma_{zz}^{tot}\rangle\sim \delta p_\alpha^2$, 
it suffices to study the 
dynamics after 
linearizing about the reference state.  Considering a contractile active fluid, 
i.e., $\Delta\mu<0$, we find that as thickness $L$ approaches  
$L_{c}$ from below, where $L_c$ is the critical thickness for the spontaneous 
flow 
instability (see Ref.~\cite{voituriez}; see also below), akin to the Frederiks 
transition in equilibrium nematics~\cite{degennes}, the Casimir stress 
$C$ diverges 
logarithmically. We find that
\begin{eqnarray}
C&=&k_BT\frac{-\pi^2}{2L_c^3}\frac{\Gamma\gamma_1}{8\eta +
\gamma_1 (\nu_1-1)^2} \ln |\frac{[2/\gamma_1 +(\nu_1 -1)^2/4\eta]
\gamma_1}{2 \Gamma\delta},\label{ctot1}
\end{eqnarray}
Here, $\Gamma =2\eta/\gamma_1 + 
(\nu_1-1)^2/4$ is a positive
dimensionless number.  The critical thickness $L_c$ is determined by the 
relation~\cite{basu}
\begin{equation}
\frac{K}{\gamma_1} \frac{\pi^2}{L_c^2} + \frac{(\nu_1-1)^2}
{4\eta}K \frac{\pi^2}{L_c^2} = -\xi\Delta\mu
\frac{\nu_1-1}{2\eta}. \label{critlength1}
\end{equation}
Clearly, 
$L_c$ diverges as $\Delta\mu \rightarrow 0$, consistent with the fact that 
there 
are no instabilities at any thickness in equilibrium.
Further we have used, 
$L=L_c(1-\delta)$, where $0<\delta \ll 1$ is a small, dimensionless number 
parameterizing the 
thickness $L$ approaching the critical thickness $L_c$ from below. Clearly, 
$C$ vanishes for 
$L_c\rightarrow \infty$ for fixed $L$ (equivalently for
$\Delta\mu =0$ for a fixed $L$), as expected.   Compare this with the 
corresponding equilibrium result
\begin{equation}
 C_{eq}= -\frac{1}{8\pi} \frac{K_BT}{L^3}\zeta_R(3), \label{ctot11}
\end{equation}
where $\zeta_R(3)$ is the Riemann-Zeta function~\cite{ajdari}. Clearly, $C_{eq}$ 
has no divergence at 
any finite $L$, in contrast to $C$ in (\ref{ctot1}).
 It follows 
from 
(\ref{ctot1}) and (\ref{ctot11}) that both $C$ and $C_{eq}$ are negative. This 
implies that the surfaces at 
$z=0$ and $z=L$ are attracted towards each other. This feature is similar to 
the equilibrium problem~\cite{ajdari}.
 Although both contributions scale as $1/L^3$, the active contribution 
clearly 
dominates the corresponding equilibrium contribution for a sufficiently small 
$\delta$. In contrast, for an extensile active system, $C$ scales as $\zeta\Delta\gamma_1\mu/(\eta L)$ for small activity, and is repulsive in nature.

{ We now argue that $C+C_{eq}=C_{tot}$  indeed has the interpretation of 
the {\em total} Casimir stress
on the system for $L<L_c$. For instance, in the contractile case consider the differences $\Delta\sigma$ in 
the total normal stresses for two different thicknesses $L_1 =
L_c(1-\delta_1)$ and $L_2=L_c (1-\delta_2)$ with $0<\delta_1,\delta_2\ll 1$. We 
note that 
\begin{equation}
 \Delta\sigma=\sigma_{zz}^{tot}|_{z=L_1}-\sigma_{zz}^{tot}|_{z=L_2}=C_{tot}|_{
z=L_1 } - C_{tot}|_{z=L_2} = C(\delta_1) + C_{eq}(L_1) - C(\delta_2)-C_{eq}(L_2).
\end{equation}
Since $\Delta\sigma$ is a measure of the change in the force per unit area on 
the wall as the thickness changes from $L_1$ to $L_2$, we can conclude that 
$C_{tot}$ can indeed be interpreted as the total Casimir stress on the system. Similar arguments can be made in the extensile case as well, with $C_{tot}=C+C_{eq}$ as the total Casimir stress.}

In order to better understand the result given by Eqs.~(\ref{ctot1}) and 
(\ref{ctot11}), we first present arguments at 
the scaling level using  a
simplified analysis of the problem that highlights the general features of the 
active contributions in (\ref{ctot1}). This is similar to the scaling analysis
of Ref.~\cite{bead}. We provide the results of the full
fluctuating hydrodynamic equations in appendix that confirm the scaling 
analysis and yield (\ref{ctot1}).

 We consider a
small perturbation to the non-flowing steady state with ${\bf p}=\hat e_z$ 
along the $z$-axis. 
In a simplified picture, we describe the
tilt of the polarity with respect to the $z$-axis normal to the film
surface by a single small angle $\theta$. The rate of variation of
the angle $\theta$  is due to the elastic nematic torque with a
Frank elastic constant $K$ and according to Eq.~(\ref{eq:pdyn}) to a
coupling to the strain rate $u$,
\begin{equation}
\label{polarization}
 \frac{\partial \theta}{\partial t}= 
 \frac{K}{\gamma_1} \nabla^2 \theta -\nu_1 u +\tilde\xi_{\perp}({\bf
 x},t).
\end{equation}
  We have added in
 this equation the thermal noise of the orientation fluctuations 
$\tilde\xi_{\perp}({\bf r},t)$
introduced above. Noise $\tilde\xi_\perp$ is  a simplified form of $\xi_{\perp 
\alpha} (t,{\bf x})$ in Eq.~(\ref{eq:pdyn}). It is Gaussian-distributed 
with zero mean and variance given by
\begin{equation}
 \langle\tilde\xi_\perp({\bf 
x},t)\tilde\xi_\perp(0,0)\rangle=2\frac{K_BT}{\gamma_1} 
\delta({\bf x})\delta (t),
\end{equation}
in analogy with (\ref{noiseII}).
We ignore  here for simplicity the tensorial character of the strain rate and 
represent it by a scalar 
$u$ which represents one of
 its typical components.

 If the polarization angle $\theta$
does not vanish, the active stress is finite and it is compensated
by the viscous stress in the film
 \begin{equation}
 \label{stress}
  \eta u \simeq \zeta \Delta \mu \theta \quad, 
 \end{equation}
where we have for simplicity ignored the noise in the stress.
Including this noise
 does not qualitatively change the final result.  The two equations 
(\ref{polarization}) and
 (\ref{stress}) can be solved by Fourier expansion both in space and time, writing
 the polarization angle as
 \begin{equation}
\theta ({\bf x},t)=\sum_n \sin (n\pi z/L)
 \int d \omega \int \frac{d{\bf q}}{(2\pi )^2} \exp i({\bf q}\cdot
 {\bf r }- \omega t)
 {\tilde  \theta} (n, \omega, {\bf q}).
 \end{equation}
  Here,
 the position vector is ${\bf x}=({\bf r}, z)$
 where ${\bf r}$ denotes the position in the plane parallel to the film,
 and the wave vector is ${\bf k}=({\bf q},n\pi/L) $
 where  ${\bf q}$ denotes the wave vector parallel to the plane, while $n$ describes
 the discrete Fourier mode along the $z$ direction.
The Fourier transform of the orientation angle satisfies the equation  
 \begin{equation}
 \label{angle}
  - i \omega  {\tilde  \theta}(n,\omega, {\bf q})= \frac{\nu_1}{\eta} \left(( \zeta \Delta \mu  -\zeta \Delta \mu_c(n))- %
  \frac{\eta K q^2}{\nu_1 \gamma_1}  \right) {\tilde \theta} +
  {\tilde \xi_{\perp}(n,\omega, {\bf q})}.
 \end{equation}
 Here, $\zeta \Delta\mu <0$ for a contractile active fluid, where as $\zeta\Delta\mu>0$ for an extensile active fluid.
 Equation~(\ref{angle}) defines the  relaxation time $\tau_n(q)$ of 
$\tilde\theta$:
\begin{equation}
 \tau_n(q)^{-1}=-\frac{\nu_1}{\eta} [\zeta \Delta \mu  -\zeta \Delta 
\mu_c(n))-
  \frac{\eta K q^2}{\nu_1 \gamma_1}  ].
  \end{equation}
    Clearly, the system gets unstable for 
$|\zeta\Delta\mu|>\zeta\Delta\mu_c$.
   We have defined here 
 $\zeta \Delta \mu_c(n)={\eta Kn^2}/{(\nu_1 \gamma_1 \pi^2L^2)}$.  We further note that in the equilibrium limit, $u=0$ in our 
simplified description and 
hence the equilibrium relaxation time $\tau_{qe,n}$ is given by
\begin{equation}
 \tau_{qe,n}^{-1}=\frac{K}{\gamma_1}\left(q^2+\frac{n^2\pi^2}{L^2}\right).
\end{equation}
  The orientation angle 
correlation function can be directly calculated form
 Eq. (\ref{angle}) leading to
 \begin{equation}
 \label{velo}
  \langle {\tilde \theta_n}({\bf q},\omega) {\tilde \theta_n}({\bf q}',\omega') 
\rangle= 
  \frac{k_BT \gamma_1
  }{\omega^2 + \tau_q^{-2}}
   (2\pi)^3 \delta({\bf q}+{\bf q}') \delta({\omega }+{\omega}').
 \end{equation}
  Using Eq.~(\ref{casidefn}), this then yields 
\begin{eqnarray}
C&=&-\frac{\pi^3k_BT}{L^3}\sum_n\int
{d^2q}\left[\frac{1}{\tau_n(q)^{-1}}-\frac{1}{\tau_{qe,n}^{-1}}\right]. \label{casihuetot}
\end{eqnarray}
Equation (\ref{casihuetot}) applies to both contractile and extensile active fluids.
 We now consider contractile and extensile active fluids separately below.

 For a contractile active fluid with $\zeta\Delta\mu <0$; clearly the system can get unstable for sufficiently large $\zeta\Delta\mu$ for a given $L$, or equivalently, for sufficiently large $L$ for a fixed $\zeta\Delta\mu$.  The nature of $C$ depends sensitively on whether $|\zeta\Delta \mu|\rightarrow\zeta\Delta\mu_c$ from below (near the the threshold for spontaneous flow instability), or $|\zeta\Delta \mu| \ll \zeta\Delta\mu_c $ (far away from the instability threshold).
  Concentrating first on the near threshold behavior of $C$, we focus only on the $n=1$ mode that is dominant near the instability threshold, which gets unstable first as $L$ approaches $L_c$ from below. For ease of notations, we  denote $\Delta\mu_c(n=1)=\Delta\mu_c$, and $\tau_{n=1}(q)^{-1}=\tau(q)^{-1},\,\tau_{qe,n=1}=\tau_{qe}$ with
 \begin{eqnarray}
 \tau(q)^{-1}&=&-\frac{\nu_1}{\eta} [\zeta \Delta \mu  -\zeta \Delta 
\mu_c)-   \frac{\eta K q^2}{\nu_1 \gamma_1}  ],\\
\tau_{qe}^{-1}&=&\frac{K}{\gamma_1}\left(q^2 + \frac{\pi^2}{L^2}\right).
 \end{eqnarray}
 If the active stress
 $\vert \zeta\Delta\mu \vert$ is larger
 than this threshold,
 the  non-flowing steady state is unstable and the film spontaneously flows. Retaining only the $n=1$ mode,
 $C$ for an orientationally ordered contractile active fluid is given by
 \begin{eqnarray}   
C&=&-\frac{ \pi^3 } { L^3 } K_BT\int { d^2q }\left(\frac{1}{ q^2+q_c^2 }- 
\frac{1}{q^2 + \frac{\pi^2}{L^2}}\right), 
\label{casihue1}
\end{eqnarray}
valid for all $|\zeta\Delta\mu |<\zeta\Delta\mu_c$.
Here, we have defined the wave vector $q_c$ such that $q_c^2=
 (\gamma_1 \nu_1/\eta)(\zeta \Delta \mu_c -
\vert  \zeta \Delta \mu \vert )/K$ and $a$ is a small length-scale cut 
off. 
Now, in the vicinity of the spontaneous flow instability, $|\zeta\Delta \mu|\rightarrow\zeta\Delta\mu_c$ from below.
Then,
\begin{equation}
C=-\frac{\pi^2k_BT}{L^3}
\ln|\frac{1+a^2q_c^2}{a^2q_c^2}|\sim -\frac{k_BT}{L^3}\ln 
|aq_c|,\label{casiheu}
\end{equation}
 retaining only the divergent contribution to $C$ as $q_c^2\rightarrow 
0$, or equivalently, $\zeta\Delta\mu\rightarrow\zeta\Delta\mu_c$ from below or 
$L\rightarrow L_{c}=\eta\kappa/(\nu_1\gamma_1 \zeta\Delta\mu_c)$ from below.
We find that the Casimir stress (\ref{casiheu}) diverges logarithmically
as $q_c\rightarrow 0$ near the instability threshold, i.e.,
$\zeta\Delta\mu \rightarrow\zeta\Delta\mu_c$. 
The Casimir stress (\ref{casiheu}) is clearly
attractive.  Comparing this with 
(\ref{ctot1}) above we note that our simplified analysis does capture the 
correct sign and the logarithmic divergence near the instability threshold. 
Compare this with the corresponding equilibrium result given in (\ref{ctot11}); clearly
has the same scaling with $L$ as $C$, but has no divergence at any finite 
$L$.

We now consider the scaling of 
$C$ far away from the threshold ($|\zeta\Delta\mu|\ll \zeta\Delta\mu_c$) as 
well. 
Assuming small 
$\zeta\Delta\mu$ (i.e., small $q_c$), we expand the denominator of
(\ref{casihuetot}) up to the linear order in $|\zeta\Delta\mu|$. We obtain for 
\begin{equation}
 C\sim-K_BT\frac{\zeta\Delta\mu\gamma_1}{\eta L}
\end{equation}
 to the leading order in $\zeta\Delta\mu$, valid for $L\ll L_c$.
Thus far away from the threshold, the leading active contribution to $C$ scales 
as $1/L$ with $L$ that is different from 
both its form near the instability threshold as well as the equilibrium 
contribution to $C$. It remains attractive, however. We thus conclude that $C$ remains attractive for 
all $L< L_c$ for a nematically ordered active fluid.

 We now discuss the extensile case, i.e., $\zeta\Delta\mu >0$ 
for which there are no instabilties at any $L$. The active Casimir stress $C$ is 
still formally defined by Eq.~(\ref{casidefn}), which yields (\ref{casihuetot}) with the sign of $\zeta\Delta\mu$ 
reversed. The
time-scale $\tau_q$ is now given by
\begin{equation}
  \tau_n(q)^{-1}=\frac{\nu_1}{\eta}\sum_n [( \zeta \Delta \mu  +\zeta \Delta 
\mu_c(n))+
  \frac{\eta K q^2}{\nu_1 \gamma_1}  ]
\end{equation}
which is positive definite implying stability. The active Casimir stress in this case 
now reads
\begin{equation}
 C=-\frac{\pi^3k_BT}{2L^3}\frac{\eta}{\gamma_1}\sum_n\int 
d^2q\left[\frac{1}{\zeta\Delta\mu 
 +\zeta\Delta\mu_c(n) + 
\eta  Kq^2/(\nu_1\gamma_1)}-\frac{1}{\tau_{qe,n}^{-1}}\right].
\end{equation}
We expand in $\Delta\mu$, assuming small activity, and extract the leading 
order active 
contribution to $C$ as
\begin{equation}
 C\sim\frac{k_BT}{L}\frac{\gamma_1}{\eta}\zeta\Delta\mu,
\end{equation}
that vanishes with $\Delta\mu$, scales with $L$ as $1/L$ and is {\em 
positive} in sign. This implies that  $C$ for an 
extensile active fluid with nematic order is repulsive to the leading order in $\zeta\Delta\mu$, in contrast to $C$ for a 
contractile active fluid, or the 
corresponding equilibrium contribution $C_E$.  Furthermore, it does not diverge for any finite $L$, 
unlike $C$ for the contractile case.

So far, we have
considered a macroscopically oriented state 
where the reference orientation is assumed to be
perpendicular to the film. An alternative choice of boundary condition would be 
a polarization  oriented parallel to the surface of the film:
$p_x=1$ as the reference state for orientation, and $p_z=0=p_y$ at $z=0,L$. 
Similar arguments show that at the scaling 
level the Casimir  stress $C$ in these conditions is 
still given by Eq.~(\ref{casiheu}). A third choice for 
boundary conditions   is
 $p_z=0$ and $p_y$ to be free at $z=0,L$ with $p_x=1$ 
as the ordered reference state. 
This is qualitatively different from what we have considered above, owing to the 
fact that
$p_y$ is a soft mode. Further, as discussed in Ref.~\cite{sumithra}, 
with this choice of the reference state there are no instabilities
at any given thickness of the system. Thus,  the Casimir stress will be significantly
different from (\ref{casiheu}). We do not discuss this case here. 

A potential biological system where the active Casimir stresses could be relevant 
is the 
thin cell cortex or the cell lamellipodium. Due to the active Casimir forces 
acting  in the direction of the thickness of the actin layer, because of the 
overall incompressibility, the active layer tends to  stretch along the in-plane 
directions. This causes the cell membrane to stretch and contributes to the {\em 
active tension} of the cell cortex. If the thickness of the system is close to 
the critical threshold of instability, the Casimir force contribution could 
become important. 

{\em Acknowledgment:} AB acknowledges partial financial support from the 
Alexander von Humboldt Stiftung (Germany) under the Research Group Linkage 
programme (2016).

\section*{Appendix}

Here, we discuss the full calculation of polarization fluctuations
in a stochastically driven active fluid layer. The scheme of the
calculations here is very similar to the detailed calculation for
the diffusion coefficient of a test particle immersed in an active
fluid layer, as given in Ref.~\cite{bead} with full details.
Nonetheless, we reproduce the basic outline here for the sake of
completeness. We start from the relations
(\ref{antisymm})-(\ref{eq:pdyn}) and determine the conjugate field
to the polarity vector $h_\alpha$ from a Frank free energy which
describes the energies of splay, bend and twist deformations by
parameters $K_1,\,K_2$ and $K_3$. For simplicity we consider here
the limit  $K_1\rightarrow\infty$ (i.e., the splay modes are
suppressed, $\nabla\cdot{\bf p}=0$). We furthermore introduce the
constraints ${\bf p}^2=1$ and $\nabla \cdot {\bf v}=0$, i.e. we
ignore fluctuations of the magnitude of $\bf p$ and we treat the
fluid as incompressible. The two constraints $\nabla\cdot {\bf p}=0$
and $p^2=1$ are imposed by two Lagrange multipliers $h_\parallel$
and $\phi$ in the free energy functional
\begin{equation}
F=\frac{1}{2}\int d^3 x[K_2 (\nabla\times {\bf p})^2 + K_3
(\partial_z {\bf p})^2 - h_\parallel p^2 + 2\phi \nabla\cdot {\bf
p}], \label{free1}
\end{equation}
where we have assumed that $\bf p$ exhibits small fluctuations
around a reference state ${\bf p}_0=\hat {\bf e}_z$, the unit vector
along the $z$-axis. The incompressibility constraint is imposed via
the pressure $P$ as Lagrange multiplier. The active fluid is
confined between two surfaces at $z=0$ and $z=L$. We impose the
following boundary conditions: no flow across the boundary surfaces
 $v_z(z=0)=0$ and $v_z(z=L)=0$ and vanishing surface shear stress at the 
boundaries: $\partial v_\alpha/\partial z=0$, at $z=0$ and $z=L$ for
$\alpha=x,y$. In addition we impose ${\bf p}(z=0)=\hat {\bf e}_z$
and ${\bf p}(z=L)=\hat {\bf e}_z$. These boundary conditions are
satisfied by the Fourier mode expansions
\begin{eqnarray}
v_\alpha({\bf x}, t)&=&\int \frac{d^2 q}{(2\pi)^2}
\frac{d\omega}{2\pi} \sum_n \tilde v_\alpha^n({\bf
q},\omega)\exp[-i\omega t + i
{\bf r\cdot q}]\cos(\frac{n\pi z}{L}),\\
v_z({\bf x}, t)&=&\int \frac{d^2 q}{(2\pi)^2} \frac{d\omega}{2\pi}
\sum_n \tilde  v_z^n({\bf q},\omega)\exp[-i\omega t + i
{\bf r\cdot q}]\sin(\frac{n\pi z}{L}),\\
p_\alpha({\bf x}, t)&=&\int \frac{d^2 q}{(2\pi)^2}
\frac{d\omega}{2\pi} \sum_n  \tilde p_\alpha^n({\bf
q},\omega)\exp[-i\omega t + i {\bf r\cdot q}]\sin(\frac{n\pi z}{L}),
\label{fourier}
\end{eqnarray}
where $\alpha=x,y$. Here, $\bf r$ is a vector in the $x-y$ plane and
the corresponding wavevector is denoted by $\bf q$. We linearize the
state of the system around a reference state with $v_\alpha=0$,
$v_z=0$ and ${\bf p} = \hat e_z$. The force balance equation
together with the incompressibility condition and the constitutive
Eq, (\ref{stresseq})  yield
 equations for the flow field
\begin{eqnarray}
-\eta(q^2 + \frac{n^2\pi^2}{L^2}) \tilde v_z^n({\bf q},t)
&=&\zeta\Delta\mu P_{z\beta}\frac{n\pi}{L}\tilde p^n_\beta +i\zeta
\Delta\mu P_{zz}q_\beta \tilde p_\beta^n
-\frac{\nu_1}{2}P_{zz}(iq_\beta\tilde h_\beta -\frac{n\pi}{L}\tilde
h_z^n) \nonumber \\ &-& \frac{\nu_1}{2}\frac{n\pi}{L}(P_{z\beta}
\tilde h_\beta^n - P_{zz}\tilde h_z^n) - \frac{1}{2} (iq_\beta
\tilde h_\beta^n -\frac{n\pi}{L} \tilde
h_z^n) + P_{z\beta }\tilde \xi_\beta^{\sigma,n} + P_{zz}\tilde \xi_z^{\sigma,n} \quad ,
\nonumber
\end{eqnarray}
\begin{eqnarray}
 -\eta(q^2 +\frac{n^2\pi^2}{L^2}) \tilde v_\alpha^n({\bf q},t) &=& \zeta\Delta\mu
 P_{\alpha\beta}\frac{n\pi}{L} \tilde p^n_\beta
+i\zeta\Delta\mu P_{\alpha z}q_\beta \tilde p_\beta^n
-\frac{\nu_1}{2}(P_{\alpha\beta}\frac{n\pi}{L} \tilde h_\beta^n
-P_{\alpha z}\frac{n\pi}{L} \tilde h_z^n) \nonumber
\\&-&
\frac{\nu_1}{2}P_{\alpha z}(iq_\beta\tilde h_\beta^n
-\frac{n\pi}{L}h_z^n)+ \frac{1}{2}\frac{n\pi}{L}\tilde h_\alpha^n +
P_{\alpha\beta} \tilde \xi_\beta^{\sigma, n} + P_{\alpha z} \tilde
\xi_z^{\sigma,n}\label{stokes2}
\end{eqnarray}
where  $\alpha,\,\beta=x$ or $y$. Here, we have introduced the
transverse projection operators $P_{zz}=q^2/(q^2  +n^2 \pi^2/L^2)$,
$P_{\alpha\beta}=\delta_{\alpha\beta} - q_\alpha q_\beta/(q^2
+n^2\pi^2/L^2)=P_{\beta\alpha}$, and $P_{\alpha z} =-i q_\alpha
(n\pi/L)/(q^2 +n^2\pi^2/L^2)=P_{z\alpha}$ and the pressure $P$ has
already been eliminated. The  noise terms $\tilde
\xi^{\sigma,n}_\alpha$ have zero-mean with variance
\begin{equation}
\langle \tilde \xi^{\sigma,n}_\alpha ({\bf q},\omega)\tilde
\xi_\beta^{\sigma,m} ({\bf q}',\omega')\rangle = 2\eta
k_BT(q^2+\frac{n^2\pi^2}{L^2}) (2\pi)^3\delta({\bf q}+{\bf
q}')\delta({\omega}+{\omega}')\delta_{\alpha\beta}\delta_{nm}
\end{equation}
where $\alpha$ and $\beta=x,y,z$.

The dynamic equation for the polarization field reads
\begin{equation}
-i\omega \tilde p^n_\alpha=-\tilde\omega^n_{\alpha
z}-\frac{K}{\gamma_1}(q^2+\frac{n^2 \pi^2}{L^2})\tilde p^n_\alpha
-\frac{1}{\gamma_1}(h_\parallel \tilde p^n_\alpha-iq_\alpha \tilde
\phi^n)-\nu_1 \tilde u^n_{\alpha z}+\tilde \xi^n_{\perp,\alpha},
\label{eqpn}
\end{equation}
with $\tilde u^n_{\alpha z}=(-\frac{n\pi}{L}\tilde v_\alpha^n +
iq_\alpha \tilde v_z^n)/2$, $\tilde \omega^n_{\alpha
z}=-(\frac{n\pi}{L}\tilde v_\alpha^n + iq_\alpha \tilde v_z^n)$ and
noise correlations
\begin{equation}
\langle\tilde \xi^n_{\perp,\alpha}({\bf q},\omega)\tilde
\xi^m_{\perp,\beta}({\bf q'},\omega')\rangle= \frac{2 K_B
T}{\gamma_1}(2\pi)^3\delta({\bf q}+{\bf
q}')\delta({\omega}+{\omega}')\delta_{\alpha\beta}\delta_{nm}.
\end{equation}
Further, with $K_2=K_3=K$ we have $h_\alpha=-\frac{\delta F}{\delta
p_\alpha} = K\nabla^2 p_\alpha +h_\parallel p_\alpha
+\nabla_\alpha\phi$ in the real space. Elimination of the Lagrange
multipliers $h_\parallel$ and $\phi$ finally leads to~\cite{bead}


\begin{eqnarray}
-\eta(q^2 +\frac{n^2\pi^2}{L^2}) \tilde v_z^n &=& P_{zz} \xi_z^{\sigma,n} + P_{z\beta}f_\beta^{\sigma,n},\\
-\eta (q^2 +\frac{n^2\pi^2}{L^2}) \tilde v_\alpha^n &=&
\zeta\Delta\mu \frac{n\pi}{L} \tilde p_\alpha^n+
\frac{\nu_1 -1}{2}\frac{n\pi}{L} K (q^2 +\frac{n^2\pi^2}{L^2}) \tilde p_\alpha^n + P_{\alpha\beta}\tilde \xi_\beta^{\sigma,n}
+P_{\alpha z}\tilde \xi_z^{\sigma, n} \quad ,\label{vi}\\
\frac{\partial \tilde p_\alpha^n}{\partial t} &=&
-\frac{K}{\gamma_1}(q^2 +\frac{n^2\pi^2}{L^2})\tilde p_\alpha^n +
\frac{\nu_1 -1}{2}\frac{n\pi}{L} \tilde v_\alpha^n +
P_{\alpha\beta}\tilde \xi_{\perp,\beta}^n + P_{\alpha z}\tilde
\xi_{\perp,z}^n \quad .\label{pi}
\end{eqnarray}
Note that $\tilde v^n_z$  decouples from $\tilde p_\alpha^n$.
Equations (\ref{vi}-\ref{pi}) may be used to obtain expressions for
the fluctuations of $\tilde p^n_\alpha$:
\begin{equation}
\left( \frac{\partial}{\partial t} + \frac{1}{\tilde\tau_q} \right)
\tilde p^n_\alpha =-\frac{n\pi}{L} \frac{\nu_1-1}{2}
\frac{P_{\alpha\beta} \tilde \xi_m^{\sigma, n} + P_{\alpha z}\tilde
\xi_z^{\sigma, n}}{\eta (q^2 + \frac{n^2\pi^2}{L^2})} +
P_{\alpha\beta} \tilde \xi_{\perp \beta}^n +P_{\alpha z}\tilde
\xi^n_{\perp,z} \quad .\label{pfinal}
\end{equation}
 where we have
identified an {effective} relaxation time   $\tilde\tau_q$ of the
polarization fluctuations $\tilde p_\alpha^n$:
\begin{equation}
\tilde\tau_q=\left[\frac{K}{\gamma_1} (q^2 +\frac{n^2\pi^2}{L^2}) +
\frac{\nu_1 -1}{2}\left(\zeta\Delta\mu +\frac{\nu_1 -1}{2}K (q^2
+\frac{n^2\pi^2}{L^2})\right) \frac{n^2\pi^2}{L^2}\frac{1}{\eta (q^2
+ \frac{n^2\pi^2}{L^2})}\right]^{-1}.
\end{equation}
For the stability of the assumed oriented state of polarization one
must have $\tilde\tau_q>0$. { Time-scale $\tilde\tau_q$ is the
analog of the time-scale $t_p(q)$ that we extract from
Eq.~(\ref{angle})}. This allows us to calculate the correlation
function of $p_\alpha^n\, (\alpha=x,y)$: We find
\begin{equation}
\langle\sigma_{zz}^e\rangle_{z=L}=-\langle (\partial_z p_i)^2\rangle_{z=L}
=-\int\frac{d^2q}{(2\pi)^2}\frac{\pi}{L} \sum_n \frac{n^2\pi^2}{L^2}
\frac{2k_BT}{\Delta_n} \left[\frac{1}{\gamma_1} +
\frac{(\nu_1-1)^2}{4\eta (q^2
+\frac{n^2\pi^2}{L^2})}\frac{n^2\pi^2}{L^2}\right],\label{cas00}
\end{equation}
where 
\begin{equation}
\Delta_n = K (q^2 +\frac{n^2\pi^2}{L^2})\left[
\frac{1}{\gamma_1} + \frac{(\nu_1 -1)^2}{4\eta (q^2
+\frac{n^2\pi^2}{L^2})}\frac{n^2\pi^2}{L^2} + \frac{\xi\Delta\mu (\nu_1 
-1)}{2\eta\kappa
(q^2 + \frac{n^2\pi^2}{L^2})^2}\frac{n^2\pi^2}{L^2}\right].
\end{equation}

Thus we obtain for the active Casimir stress in an orientationally 
ordered active fluid: Using (\ref{casidefn})
\begin{eqnarray}
C &=& -\frac{K}{2} \langle (\partial_z
p_i)^2\rangle_{z=L} + \frac{K}{2}\langle (\partial_z
p_i)^2\rangle_{z=L,\Delta\mu=0} \nonumber \\ &=&
 \frac{K}{2}\int
\frac{d^2q}{(2\pi)^2} \frac{\pi}{L} \sum_n \frac{n^2\pi^2}{L^2}
\frac{2k_BT \xi\Delta\mu(\nu_1 -1)}{2\eta K (q^2
+\frac{n^2\pi^2}{L^2})\Delta_n}.\label{cas1}
\end{eqnarray}
This holds for both contractile and extensile active fluids and
 vanishes as
$\Delta\mu$ is set to zero.

For a contractile active fluid with nematic order, $C$ diverges when $\Delta_n=0$, which
can happen with a finite $\Delta\mu <0$. The minimum
thickness for which this can happen is given by the condition
\begin{equation}
\frac{K}{\gamma_1} \frac{\pi^2}{L_c^2} + \frac{(\nu_1-1)^2}
{4\eta}K \frac{\pi^2}{L_c^2} = -\zeta\Delta\mu
\frac{\nu_1-1}{2\eta}. \label{critlength}
\end{equation}
We evaluate the active contribution in (\ref{cas1}) near the
instability threshold (for a finite $\zeta\Delta\mu <0$), i.e., as $L\rightarrow
L_{c}$ from below. In this limit, only the $n=1$ contribution diverges; the
contributions with $n>1$ are all finite. Therefore, we retain only
the $n=1$ contribution and evaluate it; we discard all higher-$n$
contributions. Define $L=L_c (1-\delta),\,\delta >0$ is a small
dimensionless number. Keeping only the divergent term contribution
as $\delta\rightarrow 0$, we obtain for the active contribution to
the Casimir stress $C$ as $L$ approaches $L_c$ from below.
\begin{equation}
C=k_BT\frac{1}{2L_c}\frac{\xi\Delta\mu(\nu_1-1)}{8\eta + \gamma_1
(\nu_1-1)^2} \ln |\frac{[2/\gamma_1 +(\nu_1 -1)^2/4\eta]
\gamma_1}{2\delta \Gamma (\nu_1-1)}|.
\end{equation}
 Substituting for $\xi\Delta\mu$ from (\ref{critlength}), we find
\begin{eqnarray}
C&=&k_BT\frac{-\pi^2}{2L_c^3}\frac{\Gamma \gamma_1}{8\eta +
\gamma_1 (\nu_1-1)^2} \ln |\frac{[2/\gamma_1 +(\nu_1 -1)^2/4\eta]
\gamma_1}{2\delta \Gamma }|,\label{ctot3}
\end{eqnarray}
same as (\ref{ctot1}) as above.
Thus, $C$ approaches 
$-\infty$ as $\delta\rightarrow 0$. Thus, it is
{\em attractive}, similar to the equilibrium
contribution~\cite{ajdari}. The equilibrium contribution may be
evaluated in straightforward ways by following Ref.~\cite{ajdari}:
One finds, at $L\rightarrow L_c$,
\begin{equation}
C_E=-\frac{1}{8\pi}\frac{k_BT}{L_c^3}\zeta_R(3).
\end{equation}
Thus, following the logic outlined in the main text, 
the total Casimir stress for an active fluid layer of thickness
$L\rightarrow L_{c}$ from below is given by
\begin{eqnarray}
C_{tot}&=&k_BT\frac{-\pi^2}{2L_c^3}\frac{\Gamma \gamma_1}{8\eta +
\gamma_1 (\nu_1-1)^2} \ln |\frac{[2/\gamma_1 +(\nu_1 -1)^2/4\eta]
\gamma_1}{2\delta \Gamma }|\nonumber
\\&-&\frac{1}{8\pi}\frac{k_BT}{L_c^3}\zeta_R(3),\label{ctot}
\end{eqnarray}
which is, of course, overall attractive. 

The scaling of the active contribution to $C$ with $L$ changes drastically for 
$L\ll L_c$. We use (\ref{cas1}) and focus on the second term on the right 
hand side of it which is the active contribution. We extract the 
${\cal O}(\zeta\Delta\mu)$ contribution for small $\zeta\Delta\mu$ that yields 
the leading order active contribution to $C$ for small $\zeta\Delta\mu$. 
We find
\begin{equation}
 C=\frac{K}{2}\int
\frac{d^2q}{(2\pi)^2} \frac{\pi}{L} \sum_n \frac{n^2\pi^2}{L^2}
\frac{2k_BT \zeta\Delta\mu(\nu_1 -1)}{2\eta K^2 (q^2
+\frac{n^2\pi^2}{L^2})^2[\frac{1}{\gamma_1} + \frac{(\nu_1-1)^2}{4\eta 
(q^2+\frac{n^2\pi^2}{L^2})}\frac{n^2\pi^2}{L^2}]}. \label{caslowL}
\end{equation}
This active contribution, being negative ($\zeta\Delta\mu<0$), remains 
attractive and clearly scales as $1/L$, different from both the equilibrium 
contribution (that scales as $1/L^3$) and the contribution for $L\rightarrow 
L_c$ from below that shows a logarithmic divergence. This is consistent with 
the predictions from our simplified analysis above.

So far, we have  considered only 
thermal noises above while averaging over the noise ensembles, keeping the 
active effects only in the deterministic parts of the dynamical model. In 
general, however, there are active noises present over and above the thermal 
noises. For simplicity, we supplement the thermal noise in (\ref{pfinal}) by an 
active noise that is assumed to be $\delta$-correlated in space and time, with a 
variance that should scale with $\Delta\mu$. The precise amplitude of the 
variance should depend on the detailed nature of the stochasticity of the motor 
movements. We now refer to Eq.~(\ref{cas00}): then to the leading order in 
$\Delta\mu$, the active noises should generate an additional active 
contribution $\delta C_A$ to $C$ in (\ref{cas1}) above near $L=L_c$. This 
is of the form
\beq
\delta C_A\sim -\frac{D_0\Delta\mu}{L_c^3}\zeta_R(3),
\eeq
where $D_0$ is a dimensional constant. 
Thus, this additional contribution is attractive, has the same scaling with $L$ 
as the equilibrium contribution $C_E$ and has no divergence as $L\rightarrow 
L_{c}$ from below. We did not consider any active, multiplicative noises that 
may 
be important in cell biology contexts as illustrated in Ref.~\cite{amit1}.

Our analyses above may be extended to obtain $C$ just above the
the threshold of the spontaneous flow instability~\cite{voituriez}. Above the 
threshold,
the steady reference state is given by $v_x=A \cos (z\pi/L),\,p_z=1,\,p_{x0}=\epsilon \sin (z\pi/L),\,
v_z=0=v_y,\,p_y=0$, with $A=4L\zeta\Delta\mu \epsilon/[\pi(4\eta + \gamma_1(\nu_1 +1)^2)]$ and
$\epsilon = \sqrt {1-L_c/L},\,L>L_c$~\cite{voituriez}. We discuss the case with 
$\epsilon\rightarrow 0$.
We  impose the same boundary conditions as above. The viscous contribution to $C_{tot}$ continues
to be zero by the same argument as above, since the spontaneous flow velocity $v_x$ has no in-plane
coordinate dependences. Defining $\delta p_x$ as the fluctuation of $p_x$ 
around $p_{x0}$, the new reference state, we note
that the boundary condition on $\delta p_x$ is same as that on $p_x$ before, i.e., for no
spontaneous flows; boundary conditions on $p_y$, having a zero value in the reference state,
naturally remains unchanged from the previous case. We, thus, conclude that $\delta p_x$ and $p_y$
follow the same (linearized) equations (\ref{pfinal}) for $p_x$ and $p_y$ as in the previous case. Hence, the solutions
for $\delta p_x$ and $p_y$ are identical to those of $p_x$ and $p_y$ in the previous case. It is now
straightforward to see that the expression for the Casimir force $C_{tot}$ as given in (\ref{ctot}) now has an additional
contribution
\begin{eqnarray}
\delta C&=& -\frac{K}{2}\langle \partial_z p_{x0}\partial_zp_{x0}\rangle 
|_{z=L} \nonumber\\
&=& - \frac{K}{2}\epsilon^2
\frac{\pi^2}{L^2}\cos^2 (\pi z/L)|_{z=L} \nonumber \\
&=& -\frac{K}{2}\frac{L-L_c}{L}\frac{\pi^2}{L^2}.\label{newcas}
\end{eqnarray}
We note that the additional contribution $\delta C_{tot}$ depends on the Frank elastic constant $K$ and has a negative sign,
displaying its attractive nature. Further and not surprisingly, it vanishes as $(L-L_c)$ as $L\rightarrow L_c$, and hence is small
just above the threshold. Thus, even above the threshold of the 
spontaneous flow instability, the dominant contribution to
$C$ still comes from (\ref{ctot}), its value just below the threshold.  
Lastly, if we continue to use the above reference states for $L_c\lesssim L$ 
even for $L\gg L_c$, then $\delta C$ scales as $1/L^2$ for $L\gg L_c$ and 
forms the dominant contribution in $C$.

 In the above we have considered a contractile active fluid. For an 
extensile system with $\xi\Delta\mu >0$, there are no divergences in 
(\ref{cas00}) or (\ref{cas1}) for any $L$. Expanding (\ref{cas1}) in 
$\zeta\Delta\mu$, we extract an active contribution linear in $\zeta\Delta\mu$ 
that 
scales with $L$ as $1/L$, different from the scaling of $C$ in the 
contractile 
case, or from the equilibrium contribution $C_E$. We find for the leading order 
active contribution to the
Casimir stress
\begin{equation}
 C=\frac{K}{2}\int
\frac{d^2q}{(2\pi)^2} \frac{\pi}{L} \sum_n \frac{n^2\pi^2}{L^2}
\frac{2k_BT \zeta\Delta\mu(\nu_1 -1)}{2\eta K^2 (q^2
+\frac{n^2\pi^2}{L^2})^2[\frac{1}{\gamma_1} + \frac{(\nu_1-1)^2}{4\eta 
(q^2+\frac{n^2\pi^2}{L^2})}\frac{n^2\pi^2}{L^2}]}\sim \frac{K_BT\zeta\Delta\mu\gamma_1}{\eta K L},
\end{equation}
that scales with $L$ as $1/L$; here only.
Thus, the active contribution comes with a  positive sign ($\zeta\Delta\mu>0$), 
i.e., 
repulsive Casimir stress, a 
feature obtained 
in our simplified analysis above. Furthermore given that $C_{eq}<0$, it is 
possible that $C_{tot}=C+C_{eq}$ changes sign as the thickness $L$ or the activity 
parameter $\zeta\Delta\mu$ is varied, potentially creating an 
intriguing crossover between a repulsive and an attractive Casimir stress.
Lastly, the differences in the active Casimir stress $C$ for the contractile 
and 
extensile cases potentially 
open up experimental routes to distinguish contractile activity from extensile 
activity by measuring $C$.

\end{document}